\documentclass{amsproc}
\usepackage{amsmath}
\usepackage{amsfonts}

\theoremstyle{plain}

\numberwithin{equation}{section}
\input{tcilatex}

\begin{document}
\title[Intuitionistic Fuzzy Ideal Extensions of $\Gamma $-Semigroups]{%
Intuitionistic Fuzzy Ideal Extensions of $\Gamma $-Semigroups}
\author{Nayyar Mehmood}
\address{School of Chemical and Material Engineering, National University of
Sciences and Technology, H-12, Islamabad.}
\email{nayyarnh@yahoo.co.uk}
\author{Imran Haider Qureshi }
\curraddr{Department of Mathematics,Quaid-i-Azam University Islamabad
Pakistan}
\email{imran\_haider707@yahoo.com}
\date{1st, Nov, 2010}

\begin{abstract}
In this paper the concept of the extensions of intuitionistic fuzzy ideals
in a semigroup has been extended to a $\Gamma $-Semigroups. Among other
results characterization of prime ideals in a $\Gamma $-Semigroups in terms
of intuitionistic fuzzy ideal extension has been obtained.
\end{abstract}

\maketitle

\section{Introduction}

$\Gamma $-Semigroups was introduced by Sen and Saha$[$\ref{9}$]$ as a
generalization of semigroup and ternary semigroup. Many results of
semigroups could be extended to $\Gamma $-Semigroups directly and via
operator semigroups$[$\ref{2}$]$ of a $\Gamma $-Semigroups.The concept of
intuitionistic fuzzy set was introduced by Atanassove$\left[ \ref{oo},\ref%
{ooo}\right] $\ ,as a generalizion of the notion of fuzzy set. Many results
of semigroups have been studied in terms of fuzzy sets$[$\ref{11}$]$. Kuroki$%
[$\ref{3},\ref{4}$]$ is the main contributor to this study. Motivated by
Kuroki $[$\ref{3},\ref{4}$]$, Xie$[$\ref{10}$]$, Mustafa et all$[$\ref{5}$]$
we have initiated the study of $\Gamma $-Semigroups in terms of
intuitionistic fuzzy sets. This paper is a continuation of $[$\ref{im}$]$,$[%
\ref{imn}]$. In this paper, the concept of the extensions of intuitionistic
fuzzy ideals in a semigroup, introduced by Xie, has been extended to the
general situation of $\Gamma $-Semigroups. We have investigated some of its
properties in terms of intuitionistic fuzzy prime and intuitionistic fuzzy
semiprime ideals of $\Gamma $-semigroup. Among other results we have
obtained characterization of prime ideals in a $\Gamma $-Semigroups in terms
of intuitionistic fuzzy ideal extension. The above introduction is mostly a
part of $[\ref{33}]$

\section{Preliminaries}

\subsection{Definition$[$\protect\ref{2}$]$}

Let $S$ and $\Gamma $ be two non-empty sets. $S$ is called a $\Gamma $%
-semigroup[\ref{2}] if there exist mappings from $S\times \Gamma \times S$
to $S$, written as $\left( a,\alpha ,b\right) $ $\rightarrow $ $a\alpha b$,
and from $\Gamma \times S\times \Gamma $ to $\Gamma $, written as $\left(
\alpha ,a,\beta \right) $ $\rightarrow $ $\alpha a\beta $ satisfying the
following associative laws%
\begin{equation*}
(a\alpha b)\beta c=a(\alpha b\beta )c=a\alpha (b\beta c)\ and\ \alpha
(a\beta b)\gamma =(\alpha a\beta )b\gamma =\alpha a(\beta b\gamma )
\end{equation*}%
for all $a,b,c\in S$ and for all $\alpha ,\beta ,\gamma \in \Gamma .$

\subsection{Definition $[$\protect\ref{oo}$]$}

An intuitionistic fuzzy set $A\ $of a non-empty set $X$ is an object having
of the form%
\begin{equation*}
\ A=\left\{ \left( x,\mu _{A}(x),\gamma _{A}(x)\right) \colon x\in X\right\}
\end{equation*}%
where the function $\mu _{A}\colon X\rightarrow \left[ 0,1\right] $ and $%
\gamma _{A}\colon X\rightarrow \left[ 0,1\right] $ denote the degree of
membership and the degree of nonmembership of each element $x\in X$ to the
set $A,$ and $0\leq \mu _{A}(x)+\gamma _{A}(x)\leq 1$ for all $x\in X.$

For the sake of simplicity, We shall use the symbol $A=(\mu _{A},\gamma
_{A}) $ \ for the intuitionistic fuzzy set $A=\left\{ \left( x,\mu
_{A}(x),\gamma _{A}(x)\right) \colon x\in S\right\} .\func{Im}\left( \mu
_{A}\right) $ denote the image set of $\mu _{A}.$Similirly $\func{Im}\left(
\gamma _{A}\right) $ denote the image set of $\gamma _{A}.$

\subsection{Definition $[$\protect\ref{10}$]$}

The set of all intuitionistic fuzzy subsets $A=(\mu _{A},\upsilon _{A})$ and 
$B=(\mu _{B},\upsilon _{B})$ of a set X with the relation%
\begin{equation*}
A\subseteq B\ \ iff\ \ \mu _{A}\leq \mu _{B}\ and\ \upsilon _{A}\geq
\upsilon _{B}
\end{equation*}
$\forall $ $x\in X$ is a complete lattice.

For a nonempty family $\left\{ A_{i}=(\mu _{A_{i}},\upsilon _{A_{i}})\ :i\in
I\right\} $ of intuitionistic fuzzy subsets of $X$, the $\inf A_{i}=\left\{
(\inf \mu _{A_{i}},\sup \upsilon _{A_{i}})\ :i\in I\right\} $ and the $\sup
A_{i}=\left\{ (\sup \mu _{A_{i}},\inf \upsilon _{A_{i}})\ :i\in I\right\} $
are the intuitionistic fuzzy subsets of $X$ defined by:

$\inf A_{i}$ $:X\longrightarrow $ $[0,1]$,$x\longrightarrow \left\{ (\inf
\mu _{A_{i}}\left( x\right) ,\sup \upsilon _{A_{i}}\left( x\right) )\ :i\in
I\right\} $

$\sup A_{i}$ $:X\longrightarrow $ $[0,1]$, x$\longrightarrow \left\{ (\sup
\mu _{A_{i}}\left( x\right) ,\inf \upsilon _{A_{i}}\left( x\right) )\ :i\in
I\right\} $ where $\inf \mu _{A_{i}}\left( x\right) =\inf \left\{ \mu
_{A_{i}}\left( x\right) :i\in I\right\} $ and $\sup \upsilon _{A_{i}}\left(
x\right) =\sup \left\{ \upsilon _{A_{i}}\left( x\right) :i\in I\right\} \ $%
and similarly for $\sup \mu _{A_{i}}\left( x\right) $ and $\inf \upsilon
_{A_{i}}\left( x\right) .$

\subsection{Definition $[$\protect\ref{imn}$]$}

A non-empty intitutionistic fuzzy subset $A=\left( \mu _{A},\upsilon
_{A}\right) $ of a $\Gamma $-semigroup S is called a intitutionistic fuzzy
left ideal(right ideal) of S if%
\begin{eqnarray*}
\mu _{A}(x\gamma y) &\geq &\mu _{A}(y)\text{ \ \ \ (}\mu _{A}(x\gamma y)\geq
\mu _{A}(y)\text{)} \\
\upsilon _{A}(x\gamma y) &\leq &\upsilon _{A}(y)\text{ \ \ \ (}\upsilon
_{A}(x\gamma y)\leq \upsilon _{A}(y)\text{)}
\end{eqnarray*}

$\forall x,y\in S,\forall \gamma \in $ $\Gamma $.

\subsection{Definition $[$\protect\ref{imn}$]$}

A non-empty intitutionistic fuzzy subset $A=\left( \mu _{A},\upsilon
_{A}\right) $ of a $\Gamma $-semigroup $S$ is called an intitutionistic
fuzzy ideal of S if it is both intitutionistic fuzzy left ideal and
intitutionistic fuzzy right ideal of $S$.

\subsection{Definition $[$\protect\ref{im}$]$}

An intitutionistic fuzzy ideal $A=\left( \mu _{A},\upsilon _{A}\right) $ of
a $\Gamma $-semigroup S is called intitutionistic fuzzy prime ideal 
\begin{equation*}
\underset{\gamma \in \Gamma }{\inf }\mu _{A}(x\gamma y)=\{\mu _{A}(x)\vee
\mu _{A}(y)\}
\end{equation*}%
and%
\begin{equation*}
\underset{\gamma \in \Gamma }{\sup }\upsilon _{A}(x\gamma y)=\{\upsilon
_{A}(x)\wedge \upsilon _{A}(y)\}
\end{equation*}%
$\forall x,y\in S$.

\subsection{Definition}

An intuitionistic fuzzy ideal $A=\left( \mu _{A},\upsilon _{A}\right) $ of a 
$\Gamma $-Semigroups $S$ is called intuitionistic fuzzy semiprime ideal if%
\begin{equation*}
\mu _{A}(x)\geq \underset{\gamma \in \Gamma }{\inf }\mu _{A}(x\gamma x)
\end{equation*}%
and%
\begin{equation*}
\underset{\gamma \in \Gamma }{\upsilon _{A}(x)\leq \sup }\upsilon
_{A}(x\gamma x)
\end{equation*}%
$\forall x,y\in S$.

\subsection{Definition $[\protect\ref{2}]$}

Let S be a $\Gamma $-Semigroups. Then an ideal $I$ of $S$ is said to be

$(i)$ prime if for ideals $A,B$ of $S$, $A\Gamma B\subseteq I$ implies that $%
A\subseteq I$ or $B\subseteq I.$

$(ii)$ semiprime if for an ideal $A$ of $S$, $A\Gamma A\subseteq I$ implies
that $A\subseteq I.$

\subsection{Proposition $[$\protect\ref{im}, \protect\ref{imn}$]$}

Let $S$ be a $\Gamma $-Semigroups and $\phi =I\subseteq S.$ Then $I$ is an
ideal(prime ideal, semiprime ideal) of $S$ iff $X=(\Phi _{I},\Psi _{I})$ is
an intuitionistic fuzzy ideal(resp. intuitionistic fuzzy prime ideal,
intuitionistic fuzzy semiprime ideal) of $S$, where $X=(\Phi _{I},\Psi _{I})$
is the characteristic function of $I$.

\subsection{Theorem $[$\protect\ref{im},\protect\ref{imn}$]$}

Let $I$ be an ideal of a $\Gamma $-Semigroups $S$. Then the following are
equivalent:

$(i)$ $I$ is prime(semiprime).

$(ii)$ for $x,y\in S$, $x\Gamma y\subseteq I$ $\implies $ $x\in I$ or $y\in
I $ (resp. $x\Gamma x\subseteq I\implies x\in I$).

$(ii)$ for $x,y\in S$, $x\Gamma S\Gamma y\subseteq I$ $\implies x\in I$ or $%
y\in I$(resp. $x\Gamma S\Gamma x\subseteq I\implies x\in I$).

\section{Intuitionistic Fuzzy Ideal Extensions}

\subsection{Definition}

Let $S$ be a $\Gamma $-Semigroups, $A=(\mu _{A},\upsilon _{A})$ be an
intuitionistic fuzzy subset of $S$ and $x\in S$, then%
\begin{equation*}
<x,\ A>\left( y\right) =\left\{ \left( y,<x,\mu _{A}>(y),<x,\upsilon
_{A}>(y)\right) \colon x\in X\right\}
\end{equation*}%
is the intuitionistic fuzzy subset of $S.$where the function $<x,\mu _{A}>:S$
$\longrightarrow $ $[0,1]$ and $<x,\upsilon _{A}>:S$ $\longrightarrow $ $%
[0,1]$ defined by $<x,\mu _{A}>(y)=\underset{\gamma \in \Gamma }{\inf }\mu
_{A}(x\gamma y)$ and $<x,\upsilon _{A}>(y)=\underset{\gamma \in \Gamma }{%
\sup }\upsilon _{A}(x\gamma y)$is called the extension of $A$ by $x.$

\textbf{Example (a):} Let $S$ be the set of all non-positive integers and $%
\Gamma $ be the set of all non-positive even integers. Then $S$ is a $\Gamma 
$-Semigroups where $a\gamma b$ and $\alpha a\beta $ denote the usual
multiplication of integers $a,\gamma ,b$ and $\alpha ,a,\beta $ respectively
with $a,b\in S$ and $\alpha ,\beta ,\gamma \in \Gamma $. Let $A=(\mu
_{A},\gamma _{A})$ be an intuitionistic fuzzy subset of $S$, defined as
follows%
\begin{equation*}
\mu _{A}\left( x\right) =\left\{ 
\begin{array}{c}
1\ \ \ \ \ \ \ \ \ \ \ \ \ \ \ \ \ \ \text{if \ }x=0\ \ \ \ \  \\ 
0.1\ \ \ \ \ \ \ \ \ \ \ \ \ \ \ \ \ \text{if \ }x=-1,-2 \\ 
0.2\ \ \ \ \ \ \ \ \ \ \ \ \ \ \ \ \ \text{if \ }x<-2\ \ \ \ \ \ 
\end{array}%
\right.
\end{equation*}%
and%
\begin{equation*}
\upsilon _{A}\left( x\right) =\left\{ 
\begin{array}{c}
0\ \ \ \ \ \ \ \ \ \ \ \ \ \text{if \ }x=0 \\ 
0.7\ \ \ \ \ \ \ \ \ \text{if \ }x<0%
\end{array}%
\right.
\end{equation*}%
Then the intuitionistic fuzzy subset $A=(\mu _{A},\gamma _{A})$ of $S$ is an
intuitionistic fuzzy ideal of $S.$

For $x=0$ $\in S$, $<x,\mu _{A}>(y)=1\ $and $<x,\upsilon _{A}>(y)=0\ \forall
y\in S$. For all other $x\in S$, $<x,\mu _{A}>(y)$ $=0.2$ and $<x,\upsilon
_{A}>(y)=0.7\ \forall y\in S.$

Thus $<x,A>$ is an intuitionistic fuzzy ideal extension of $A$ by $x.$

\subsection{Proposition}

Let $A=(\mu _{A},\gamma _{A})$ be an intuitionistic fuzzy ideal of a
commutative $\Gamma $-Semigroups $S$ and $x\in S$. Then $<x,A>$ is an
intuitionistic fuzzy ideal of $S.$

\begin{proof}
Let $A=(\mu _{A},\gamma _{A})$ be an intuitionistic fuzzy ideal of a
commutative $\Gamma $-Semigroups $S$ and $p,q\in S$, $\beta $ $\in $ $\Gamma 
$. Then 
\begin{equation*}
<x,\mu _{A}>(p\beta q)=\underset{\gamma \in \Gamma }{\inf }%
%TCIMACRO{\U{3bc} }%
%BeginExpansion
\mu
%EndExpansion
_{A}(x\gamma p\beta q)\geq \underset{\gamma \in \Gamma }{\inf }%
%TCIMACRO{\U{3bc} }%
%BeginExpansion
\mu
%EndExpansion
_{A}(x\gamma p)=<x,\mu _{A}>(p)
\end{equation*}%
and%
\begin{equation*}
<x,\upsilon _{A}>(p\beta q)=\underset{\gamma \in \Gamma }{\sup }\upsilon
_{A}(x\gamma p\beta q)\leq \underset{\gamma \in \Gamma }{\sup }\upsilon
_{A}(x\gamma p)=<x,\upsilon _{A}>(p)
\end{equation*}%
Thus $<x,A>$ is an intuitionistic fuzzy right ideal of S. Hence S being
commutative $<x,A>$ is an intuitionistic fuzzy ideal of $S.$
\end{proof}

\subsection{Remark}

Commutativity of $\Gamma $-Semigroups $S$ is not required to prove that $%
<x,A>$ is an intuitionistic fuzzy right ideal of $S$ when $A=(\mu
_{A},\gamma _{A})$ is an intuitionistic fuzzy right ideal of $S$.

\subsection{Proposition}

Let $S$ be a commutative $\Gamma $-Semigroups and $A=(\mu _{A},\upsilon
_{A}) $ be an intuitionistic fuzzy prime ideal of $S$. Then $<x,A>$ is
intuitionistic fuzzy prime ideal of $S$ for all $x\in S$.

\begin{proof}
Let $A=(\mu _{A},\upsilon _{A})$ be an intuitionistic fuzzy prime ideal of
S. Then by Proposition 3.2, $<x,A>$ is an intuitionistic fuzzy ideal of $S$.
Let $y,z\in S$. Then%
\begin{eqnarray*}
\underset{\beta \in \Gamma }{\inf } &<&x,\mu _{A}>(y\beta z)=\underset{\beta
\in \Gamma }{\inf }\underset{\gamma \in \Gamma }{\inf }%
%TCIMACRO{\U{3bc} }%
%BeginExpansion
\mu
%EndExpansion
_{A}(x\gamma y\beta z)\ \ by\ \ 3.1 \\
&=&\underset{\beta \in \Gamma }{\inf }\{\mu _{A}(x)\vee 
%TCIMACRO{\U{3bc} }%
%BeginExpansion
\mu
%EndExpansion
_{A}(y\beta z)\}\ \ by\ \ 2.6 \\
&=&\{\mu _{A}(x)\vee \underset{\beta \in \Gamma }{\inf }%
%TCIMACRO{\U{3bc} }%
%BeginExpansion
\mu
%EndExpansion
_{A}(y\beta z)\} \\
&=&\{\mu _{A}(x)\vee \{%
%TCIMACRO{\U{3bc} }%
%BeginExpansion
\mu
%EndExpansion
_{A}(y)\vee 
%TCIMACRO{\U{3bc} }%
%BeginExpansion
\mu
%EndExpansion
_{A}(z)\} \\
&=&\left\{ (\mu _{A}(x)\vee 
%TCIMACRO{\U{3bc} }%
%BeginExpansion
\mu
%EndExpansion
_{A}(y))\vee \left( \mu _{A}(x)\vee 
%TCIMACRO{\U{3bc} }%
%BeginExpansion
\mu
%EndExpansion
_{A}(z)\right) \right\} \\
&=&\left\{ \underset{\ell \in \Gamma }{\inf }%
%TCIMACRO{\U{3bc} }%
%BeginExpansion
\mu
%EndExpansion
_{A}(x\ell y)\vee \underset{\varepsilon \in \Gamma }{\inf }%
%TCIMACRO{\U{3bc} }%
%BeginExpansion
\mu
%EndExpansion
_{A}(x\varepsilon z)\right\} \\
&=&<x,\mu _{A}>\left( y\right) \vee <x,\mu _{A}>\left( z\right)
\end{eqnarray*}%
and%
\begin{eqnarray*}
\underset{\beta \in \Gamma }{\sup } &<&x,\upsilon _{A}>(y\beta z)=\underset{%
\beta \in \Gamma }{\sup }\underset{\gamma \in \Gamma }{\sup }\upsilon
_{A}(x\gamma y\beta z)\ \ by\ \ 3.1 \\
&=&\underset{\beta \in \Gamma }{\sup }\{\upsilon _{A}(x)\wedge \upsilon
_{A}(y\beta z)\}\ \ by\ \ 2.6 \\
&=&\{\upsilon _{A}(x)\wedge \underset{\beta \in \Gamma }{\sup }\upsilon
_{A}(y\beta z)\} \\
&=&\{\upsilon _{A}(x)\wedge \{\upsilon _{A}(y)\wedge \upsilon _{A}(z)\} \\
&=&\left\{ (\upsilon _{A}(x)\wedge \upsilon _{A}(y))\wedge \left( \upsilon
_{A}(x)\wedge \upsilon _{A}(z)\right) \right\} \\
&=&\left\{ \underset{\ell \in \Gamma }{\sup }\upsilon _{A}(x\ell y)\wedge 
\underset{\varepsilon \in \Gamma }{\sup }\upsilon _{A}(x\varepsilon
z)\right\} \\
&=&<x,\upsilon _{A}>\left( y\right) \wedge <x,\upsilon _{A}>\left( z\right)
\end{eqnarray*}%
Hence by Definition 2.6, $<x,A>$ is an intuitionistic fuzzy prime ideal of $%
S $.
\end{proof}

\subsection{Definition}

Suppose $S$ is a $\Gamma $-Semigroups and $A=(\mu _{A},\upsilon _{A})$ is an
intuitionistic fuzzy subset of $S$. Then we define $supp\ \mu _{A}=\{x\in
S:\mu _{A}(x)>0\}\ $and $inff\ \upsilon _{A}=\{x\in S:\upsilon _{A}(x)<1\}$

\subsection{Proposition}

Let $S$ be a $\Gamma $-Semigroups, $A=(\mu _{A},\upsilon _{A})$ be an
intuitionistic fuzzy ideal of $S$ and $x\in S$. Then we have the following:

$(1)$ $A\subseteq <x,A>$ .

$(2)$ $<(x\alpha )^{n}x,A>\subseteq <(x\alpha )^{n+!}x,A>$ $\forall \alpha $ 
$\in $ $\Gamma $, $\forall n\in N$.

$(3)If\ \mu _{A}(x)>0$ \ and $\upsilon _{A}(x)<1\ $then $supp<x,\mu _{A}>=S$
and $inff<x,\upsilon _{A}>=S.$

\begin{proof}
$\left( 1\right) .$Let $y\in S$. Then%
\begin{equation*}
<x,\mu _{A}>(y)=\underset{\gamma \in \Gamma }{\inf }\mu _{A}(x\gamma y)\geq
\mu _{A}(y)
\end{equation*}%
and%
\begin{equation*}
<x,\upsilon _{A}>(y)=\underset{\gamma \in \Gamma }{\sup }\upsilon
_{A}(x\gamma y)\leq \upsilon _{A}(y)
\end{equation*}%
(since $A$ is an intuitionistic fuzzy ideal of $S$).Hence $A\subseteq <x,A>.$

$\left( 2\right) .$In $\left( 2\right) $ we have to prove that 
\begin{eqnarray*}
( &<&(x\alpha )^{n}x,\mu _{A}>)\leq (<(x\alpha )^{n+!}x,\mu _{A})\ and\  \\
( &<&(x\alpha )^{n}x,\upsilon _{A}>)\geq (<(x\alpha )^{n+!}x,\upsilon _{A}>)
\end{eqnarray*}%
Now%
\begin{equation*}
(<(x\alpha )^{n+1}x,\mu _{A}>)\left( y\right) =\underset{\gamma \in \Gamma }{%
\inf }\mu _{A}\left( (x\alpha )^{n+1}x\gamma y\right)
\end{equation*}%
\begin{eqnarray*}
\ \ \ \ \ \ \ \ \ \ \ \ \ \ \ \ \ \ \ \ \ \ \ \ \ \ \ \ \ \ \ \ \ \ \ &=&%
\underset{\gamma \in \Gamma }{\inf }\mu _{A}\left( x\alpha \left( x\alpha
\right) ^{n}x\gamma y\right) \\
&\geq &\underset{\gamma \in \Gamma }{\inf }\mu _{A}\left( \left( x\alpha
\right) ^{n}x\gamma y\right) \\
&=&<(x\alpha )^{n}x,\mu _{A}>\left( y\right)
\end{eqnarray*}%
and%
\begin{equation*}
(<(x\alpha )^{n+1}x,\upsilon _{A}>)\left( y\right) =\underset{\gamma \in
\Gamma }{\sup }\upsilon _{A}\left( (x\alpha )^{n+1}x\gamma y\right)
\end{equation*}%
\begin{eqnarray*}
\ \ \ \ \ \ \ \ \ \ \ \ \ \ \ \ \ \ \ \ \ \ \ \ \ \ \ \ \ \ \ \ \ \ \ &=&%
\underset{\gamma \in \Gamma }{\sup }\upsilon _{A}\left( x\alpha \left(
x\alpha \right) ^{n}x\gamma y\right) \\
&\leq &\underset{\gamma \in \Gamma }{\sup }\upsilon _{A}\left( \left(
x\alpha \right) ^{n}x\gamma y\right) \\
&=&<(x\alpha )^{n}x,\upsilon _{A}>\left( y\right)
\end{eqnarray*}%
Hence $<(x\alpha )^{n}x,A>\subseteq <(x\alpha )^{n+!}x,A>$ $\forall \alpha $ 
$\in $ $\Gamma $, $\forall n\in N$.

$\left( 3\right) .$Since $<x,A>$ is an intuitionistic fuzzy subset of S, by
definition, supp $<x,A>\subseteq S.$Let $y\in S$. Since $A$ is an
intuitionistic fuzzy ideal of $S$, we have,%
\begin{equation*}
<x,\mu _{A}>(y)=\underset{\gamma \in \Gamma }{\inf }\mu _{A}(x\gamma y)\geq
\mu _{A}(x)>0
\end{equation*}%
and%
\begin{equation*}
<x,\upsilon _{A}>(y)=\underset{\gamma \in \Gamma }{\sup }\upsilon
_{A}(x\gamma y)\leq \upsilon _{A}(x)<1
\end{equation*}%
Then $<x,\mu _{A}>(y)>0\ $and $<x,\upsilon _{A}>(y)<1.$So $y\in supp<x,\mu
_{A}>$ and $y\in inff<x,\upsilon _{A}>$.
\end{proof}

\subsection{Remark}

If we consider $(x\alpha )^{0}x=x$ then $(2)$ is also true for $n=0.$

\subsection{Definition}

Suppose $S$ is a $\Gamma $-Semigroups, $M\subseteq S$ and $x\in S$. We
define $<x,M>=\{y\in S|x\Gamma y\subseteq M\}$, where $x\Gamma y:=\{x\alpha
y:\alpha \in \Gamma \}$.

\subsection{Proposition}

Let be a $\Gamma $-Semigroups and $\phi =M\subseteq S$. Then $<x,\Phi
_{M}>=\Phi _{<x,M>}$ and $<x,\Psi _{M}>=\Psi _{<x,M>}$ for every $x\in S$,
where $(\Phi _{M},\Psi _{M})$ denotes the characteristic function of $M$,
where%
\begin{equation*}
\Phi _{M}(x)=\left\{ 
\begin{array}{c}
1\text{ }if\text{ }x\in M \\ 
0\text{ }if\text{ }x\notin M%
\end{array}%
\right. \ \ ,\ \ \Psi _{M}\left( x\right) =\left\{ 
\begin{array}{c}
0\text{ }if\text{ }x\in M \\ 
1\text{ }if\text{ }x\notin M%
\end{array}%
\right.
\end{equation*}

\begin{proof}
Let $x,y\in S$.Then two cases may arise viz. Case $(i)$ $y\in <x,M>.$Case $%
(ii)$ $y\not\in <x,M>$ .

Case $(i)$ $y\in <x,M>.$Then $x\Gamma y\subseteq M$. Hence $x\gamma y\in M$ $%
\forall \gamma \in \Gamma $. This means $\Phi _{M}\left( x\gamma y\right) =1$
and $\Psi _{M}\left( x\gamma y\right) =0$ $\forall \gamma \in \Gamma $.
Hence $\underset{\gamma \in \Gamma }{\inf }\Phi _{M}\left( x\gamma y\right)
=1\ $and $\underset{\gamma \in \Gamma }{\sup }\Psi _{M}\left( x\gamma
y\right) =0$ whence $<x,\Phi _{M}>=1$ and $<x,\Psi _{M}>=0.$ Also $\Phi
_{<x,M>}=1$ and $\Psi _{<x,M>}=0.$

Case $(ii)$ $y\not\in <x,M>$.Then there exists $\gamma $ $\in $ $\Gamma $
such that $x\gamma y\not\in M.$So $\Phi _{M}\left( x\gamma y\right) =0$ and $%
\Psi _{M}\left( x\gamma y\right) =1.$Hence $\underset{\gamma \in \Gamma }{%
\inf }\Phi _{M}\left( x\gamma y\right) =0\ $and $\underset{\gamma \in \Gamma 
}{\sup }\Psi _{M}\left( x\gamma y\right) =1.$Thus $<x,\Phi _{M}>=0$ and $%
<x,\Psi _{M}>=1.$Again $\Phi _{<x,M>}=0$ and $\Psi _{<x,M>}=1.$Thus we
conclude $<x,\Phi _{M}>=\Phi _{<x,M>}$ and $<x,\Psi _{M}>=\Psi _{<x,M>}.$
\end{proof}

\subsection{Proposition}

Let $S$ be a $\Gamma $-Semigroups and $A=(\mu _{A},\upsilon _{A})$ be a
nonempty intuitionistic fuzzy subset of $S$. Then for any $t\in \left[ 0,1%
\right] $, $<x,A_{t}>=<x,A>_{t}$for all $x\in S.$where $A_{t}$ denotes $%
U(\mu _{A}\colon t)$ and $L(\upsilon _{A}\colon t).$

\begin{proof}
Let $y\in <x,A>_{t}.$This means $y\in U(<x,\mu _{A}>\colon t)$ and $y\in
L(<x,\upsilon _{A}>\colon t).$Then $<x,\mu _{A}>(y)\geq t$ and $<x,\upsilon
_{A}>(y)\leq t$.Hence $\underset{\gamma \in \Gamma }{\inf }\mu _{A}(x\gamma
y)\geq t$ and $\underset{\gamma \in \Gamma }{\sup }\upsilon _{A}(x\gamma
y)\leq t.$This gives $\mu _{A}(x\gamma y)\geq t\ $and $\upsilon _{A}(x\gamma
y)\leq t$ for all $\gamma $ $\in $ $\Gamma $ and hence $x\gamma y\in $ $%
U(\mu _{A}\colon t)$ and $x\gamma y\in $ $L(\upsilon _{A}\colon t).$for all $%
\gamma \in \Gamma $. Consequently,$y\in <x,U(\mu _{A}\colon t)>$ and $y\in
<x,L(\mu _{A}\colon t)>.$i.e $y\in <x,A_{t}>.$It follows that $%
<x,A>_{t}\subseteq <x,A_{t}>.$Reversing the above argument we can deduce
that $<x,A_{t}>\subseteq <x,A>_{t}.$Hence $<x,A_{t}>=<x,A>_{t}.$
\end{proof}

\subsection{Proposition}

Let $S$ be a commutative $\Gamma $-Semigroups i.e., $a\alpha b=b\alpha a$ $%
\forall a,b\in S$, $\forall \alpha \in \Gamma $ and $A=(\mu _{A},\upsilon
_{A})$ be an intuitionistic fuzzy subset of $S$ such that $<x,A>=$ $A$ for
every $x\in S.$Then $A=(\mu _{A},\upsilon _{A})$ is a constant function.

\begin{proof}
Let $x,y\in S$. Then by hypothesis we have%
\begin{eqnarray*}
\mu _{A}(x) &=&<y,%
%TCIMACRO{\U{3bc} }%
%BeginExpansion
\mu
%EndExpansion
_{A}>(x) \\
&=&\underset{\gamma \in \Gamma }{\inf }\mu _{A}(y\gamma x) \\
&=&\underset{\gamma \in \Gamma }{\inf }\mu _{A}(x\gamma y) \\
&=&<x,\mu _{A}>(y)=\mu _{A}(y)
\end{eqnarray*}%
and%
\begin{eqnarray*}
\upsilon _{A}(x) &=&<y,\upsilon _{A}>(x) \\
&=&\underset{\gamma \in \Gamma }{\sup }\upsilon _{A}(y\gamma x) \\
&=&\underset{\gamma \in \Gamma }{\sup }\upsilon _{A}(x\gamma y) \\
&=&<x,\upsilon _{A}>(y)=\upsilon _{A}(y)
\end{eqnarray*}%
Hence $A=(\mu _{A},\upsilon _{A})$ is a constant function.
\end{proof}

\subsubsection{Corollary}

Let $S$ be a commutative $\Gamma $-Semigroups, $A=(\mu _{A},\upsilon _{A})$
be an intuitionistic fuzzy prime ideal of $S$. If $A=(\mu _{A},\upsilon
_{A}) $ is not constant, then $A=(\mu _{A},\upsilon _{A})$ is not a maximal
intuitionistic fuzzy prime ideal of $S$.

\begin{proof}
Let $A=(\mu _{A},\upsilon _{A})$ be an intuitionistic fuzzy prime ideal of $%
S $. Then, by Proposition 3.4 for each $x\in S$, $<x,A>$ is an
intuitionistic fuzzy prime ideal of $S.$Now by Proposition $3.6\ (1)$ $%
A\subseteq <x,A>$ for all $x\in S.$If $<x,A>=$ $A$ for all $x\in S$ then by
Proposition $3.11$ $A$ is constant which is not the case by hypothesis.Hence
there exists $x\in S$ such that $A\subset <x,A>.$This completes the proof.
\end{proof}

\subsection{Proposition}

Let $S$ be a commutative $\Gamma $-Semigroups. If $A=(\mu _{A},\upsilon
_{A}) $ is an intuitionistic fuzzy semiprime ideal of $S$, then $<x,A>$ is
an intuitionistic fuzzy semiprime ideal of $S$ for every $x\in S$.

\begin{proof}
Let $A=(\mu _{A},\upsilon _{A})$ be an intuitionistic fuzzy semiprime ideal
of $S$ and $x,y\in S$. Then $\underset{\gamma \in \Gamma }{\inf }<x,\mu
_{A}>(y\gamma y)=\underset{\gamma \in \Gamma }{\inf }\underset{\delta \in
\Gamma }{\inf }%
%TCIMACRO{\U{3bc} }%
%BeginExpansion
\mu
%EndExpansion
_{A}(x\delta y\gamma y)$ $\leq \underset{\gamma \in \Gamma }{\inf }\underset{%
\delta \in \Gamma }{\inf }%
%TCIMACRO{\U{3bc} }%
%BeginExpansion
\mu
%EndExpansion
_{A}(x\delta y\gamma y\delta x)$ (since $A$ is an intuitionistic fuzzy ideal
of $S$) $=\underset{\gamma \in \Gamma }{\inf }\underset{\delta \in \Gamma }{%
\inf }%
%TCIMACRO{\U{3bc} }%
%BeginExpansion
\mu
%EndExpansion
_{A}(x\delta y\gamma x\delta y)$ (using commutativity of $S$ and Definition $%
2.7$) $=<x,\mu _{A}>\left( y\right) .$ $\underset{\gamma \in \Gamma }{\text{%
And }\sup }<x,\upsilon _{A}>(y\gamma y)=\underset{\gamma \in \Gamma }{\sup }%
\underset{\delta \in \Gamma }{\sup }\upsilon _{A}(x\delta y\gamma y)$ $\geq 
\underset{\gamma \in \Gamma }{\sup }\underset{\delta \in \Gamma }{\sup }%
\upsilon _{A}(x\delta y\gamma y\delta x)$ (since $A$ is an intuitionistic
fuzzy ideal of $S$) $=\underset{\gamma \in \Gamma }{\sup }\underset{\delta
\in \Gamma }{\sup }\upsilon _{A}(x\delta y\gamma x\delta y)$ (using
commutativity of $S$ and Definition $2.7$) $=<x,\upsilon _{A}>\left(
y\right) .$

Again by Proposition $3.2$, $<x,A>$ is an intuitionistic fuzzy ideal of $S$.
Consequently,$<x,A>$ is an intuitionistic fuzzy semiprime ideal of S for all 
$x\in S$.
\end{proof}

\subsection{Corollary}

Let $S$ be a commutative $\Gamma $-Semigroups, $\{A_{i}\}_{i\in I}$ be a
non-empty family of intuitionistic fuzzy semiprime ideals of $S$ and let $%
A=(\mu _{A},\upsilon _{A})$ $=\left( \inf \mu _{A_{i}},\sup \upsilon
_{A_{i}}\right) _{i\in I}.$Then for any $x\in S$, $<x,A>$ is an
intuitionistic fuzzy semiprime ideal of $S$.

\begin{proof}
Since each $A_{i}=(\mu _{A_{i}},\upsilon _{A_{i}})$ $\left( i\in I\right) $
is an intuitionistic fuzzy ideal, $\mu _{A_{i}}\left( 0\right) \not=0$ and $%
\upsilon _{A_{i}}\left( 0\right) \not=1$ $\forall i\in I$ $($Each $\mu
_{A_{i}}$and $\upsilon _{A_{i}}$ are non-empty, so there exists $xi\in S$
such that $\mu _{A_{i}}(x_{i})\not=0$ and $\upsilon _{A_{i}}(x_{i})\not=1$ $%
\forall i\in I$. Also $\mu _{A_{i}}(0)=\mu _{A_{i}}(0\gamma x_{i})\geq \mu
_{A_{i}}(x_{i})$ and $\upsilon _{A_{i}}(0)=\upsilon _{A_{i}}(0\gamma
x_{i})\leq \upsilon _{A_{i}}(x_{i})$ $\forall i\in I.$Hence $\forall i\in
I,\mu _{A_{i}}(0)\not=0$ and $\upsilon _{A_{i}}\left( 0\right) \not=1).$%
Consequently, $%
%TCIMACRO{\U{3bc} }%
%BeginExpansion
\mu
%EndExpansion
_{A}\not=0$ and $\upsilon _{A}\not=1.$Thus $A$ is non-empty.Now let $x,y\in
S $. Then%
\begin{eqnarray*}
\mu _{A}(x\gamma y) &=&\inf \left\{ \mu _{A_{i}}:i\in I\right\} (x\gamma y)
\\
&=&\inf \left\{ \mu _{A_{i}}(x\gamma y):i\in I\right\} \\
&\geq &\inf \left\{ \mu _{A_{i}}(x):i\in I\right\} \\
&=&\mu _{A}(x)
\end{eqnarray*}%
and 
\begin{eqnarray*}
\upsilon _{A}(x\gamma y) &=&\sup \left\{ \upsilon _{A_{i}}:i\in I\right\}
(x\gamma y) \\
&=&\sup \left\{ \upsilon _{A_{i}}(x\gamma y):i\in I\right\} \\
&\leq &\sup \left\{ \upsilon _{A_{i}}(x):i\in I\right\} \\
&=&\upsilon _{A}(x)
\end{eqnarray*}%
Hence $S$ being a commutative $\Gamma $-Semigroups $A$ is an intuitionistic
fuzzy ideal of $S$.

Now if $a\in S$ then 
\begin{eqnarray*}
\mu _{A}(a) &=&\inf \left\{ \mu _{A_{i}}:i\in I\right\} (a) \\
&=&\inf \left\{ \mu _{A_{i}}(a):i\in I\right\} \\
&\geq &\inf \left\{ \underset{\gamma \in \Gamma }{\inf }\mu _{A_{i}}(a\gamma
a):i\in I\right\} \ cf.Definition2.7 \\
&=&\underset{\gamma \in \Gamma }{\inf }\{\inf \left\{ \mu _{A_{i}}(a\gamma
a):i\in I\right\} \\
&=&\underset{\gamma \in \Gamma }{\inf }\{\inf \left\{ \mu _{A_{i}}:i\in
I\right\} (a\gamma a) \\
&=&\underset{\gamma \in \Gamma }{\inf }\mu _{A}(a\gamma a)
\end{eqnarray*}%
and%
\begin{eqnarray*}
\upsilon _{A}(a) &=&\sup \left\{ \upsilon _{A_{i}}:i\in I\right\} (a) \\
&=&\sup \left\{ \upsilon _{A_{i}}(a):i\in I\right\} \\
&\leq &\sup \left\{ \underset{\gamma \in \Gamma }{\sup }\upsilon
_{A_{i}}(a\gamma a):i\in I\right\} \ cf.Definition2.7 \\
&=&\underset{\gamma \in \Gamma }{\sup }\{\sup \left\{ \upsilon
_{A_{i}}(a\gamma a):i\in I\right\} \\
&=&\underset{\gamma \in \Gamma }{\sup }\{\sup \left\{ \upsilon _{A_{i}}:i\in
I\right\} (a\gamma a) \\
&=&\underset{\gamma \in \Gamma }{\sup }\upsilon _{A}(a\gamma a)
\end{eqnarray*}%
This means,$A=(\mu _{A},\upsilon _{A})$ is an intuitionistic fuzzy semiprime
ideal of $S$. Hence by Proposition $3.13$, for any $x\in S$ $<x,A>$ is an
intuitionistic fuzzy semiprime ideal of $S$.
\end{proof}

\subsection{Remark}

The proof of the above Corollary shows that in a $\Gamma $-Semigroups
intersection of arbitrary family of intuitionistic fuzzy semiprime ideals is
an intuitionistic fuzzy semiprime ideal.

\subsection{Corollary}

Let $S$ be a commutative $\Gamma $-Semigroups, $\{S_{i}\}_{i\in I}$ a
non-empty family of semiprime ideals of $S$ and $A:=\cap _{i\in
I}S_{i}\not=\phi .$ Then $<x,X_{A}>$ is an intuitionistic fuzzy semiprime
ideal of $S$ for all $x\in S$ where $X_{A}=(\Phi _{A},\Psi _{A})$ is the
characteristic function of A.

\begin{proof}
By supposition $A=\phi .$ Then for any ideal $P$ of $S$, $P\Gamma P\subseteq
A$ implies that $P\Gamma P\subseteq S_{i}$ $\forall i\in I$. Since each $%
S_{i}$ is a semiprime ideal of $S$, $P\subseteq S_{i}\ \forall i\in I\ $(cf.
Definition $2.8$). So $P\subseteq \cap _{i\in I}S_{i}=A.$ Hence $A$ is a
semiprime ideal of $S$(cf.Definition 2.8). So the characteristic function $%
X_{A}=(\Phi _{A},\Psi _{A})$ of $A$ is an intuitionistic fuzzy semiprime
ideal of S(cf. Proposition $2.9$). Hence by Proposition $3.13$,$\forall x\in
S$ $<x,X_{A}>$ is an intuitionistic fuzzy semiprime ideal of $S.$

\textbf{Alternative Proof:} $A:=\cap _{i\in I}S_{i}\not=\phi \ $(by the
given condition). Hence $X_{A}\not=\phi \ $i.e $\Phi _{A}\not=\phi \ and\
\Psi _{A}\not=\phi \ .$Let $x\in S$. Then $x\in A$ or $x\not\in A$. If $x\in
A$ then $\Phi _{A}\left( x\right) =1$ and $\Psi _{A}\left( x\right) =0$ and $%
x\in S_{i}$ $\forall i\in I$. Hence 
\begin{equation*}
\inf \left\{ \Phi _{S_{i}}:i\in I\right\} \left( x\right) =\inf \left\{ \Phi
_{S_{i}}\left( x\right) :i\in I\right\} =1=\Phi _{A}\left( x\right)
\end{equation*}%
and%
\begin{equation*}
\sup \left\{ \Psi _{S_{i}}:i\in I\right\} \left( x\right) =\sup \left\{ \Psi
_{S_{i}}\left( x\right) :i\in I\right\} =0=\Psi _{A}\left( x\right)
\end{equation*}%
If $x\not\in A$ then $\Phi _{A}\left( x\right) =0$ and $\Psi _{A}\left(
x\right) =1$and for some $i\in I$, $x\not\in S_{i}.$ It follows that $\Phi
_{S_{i}}(x)=0$ and $\Psi _{S_{i}}\left( x\right) =1.$Hence 
\begin{equation*}
\inf \left\{ \Phi _{S_{i}}:i\in I\right\} \left( x\right) =\inf \left\{ \Phi
_{S_{i}}\left( x\right) :i\in I\right\} =0=\Phi _{A}\left( x\right)
\end{equation*}%
and%
\begin{equation*}
\sup \left\{ \Psi _{S_{i}}:i\in I\right\} \left( x\right) =\sup \left\{ \Psi
_{S_{i}}\left( x\right) :i\in I\right\} =1=\Psi _{A}\left( x\right)
\end{equation*}%
Thus we see that $\Phi _{A}=\inf \left\{ \Phi _{S_{i}}:i\in I\right\} $ and $%
\Psi _{A}=\sup \left\{ \Psi _{S_{i}}:i\in I\right\} .$Again $%
X_{S_{i}}=\left( \Phi _{S_{i}},\Psi _{S_{i}}\right) $ is an intuitionistic
fuzzy semiprime ideal of $S$ for all $i\in I\left( \text{cf.Definition }%
2.9\right) .$Consequently by Corollary 3.14, for all $x\in S$,$<x,X_{A}>$ is
an intuitionistic fuzzy semiprime ideal of $S$.
\end{proof}

\subsection{Theorem}

Let $S$ be a $\Gamma $-Semigroups. If $A=(\mu _{A},\upsilon _{A})$ is an
intuitionistic fuzzy prime ideal of $S$ and $x\in S$ such that $A(x)=\left( 
\underset{y\in S}{\inf }\mu _{A}(y),\underset{y\in S}{\sup }\upsilon
_{A}(y)\right) ,$then $<x,A>=$ $A.$Conversely, if $A=(\mu _{A},\upsilon
_{A}) $ is an intuitionistic fuzzy ideal of $S$ such that $<y,A>=A$ $\forall
y\in S $ with $A(y)$ not maximal in $A(S)$ then $A=(\mu _{A},\upsilon _{A})$
is prime.

\begin{proof}
Let $A=(\mu _{A},\upsilon _{A})$ be an intuitionistic fuzzy prime ideal of S
and $x\in S$ be such that $\mu _{A}(x)$ $=\underset{y\in S}{\inf }\mu
_{A}(y) $ and $\upsilon _{A}(x)=\underset{y\in S}{\sup }\upsilon _{A}(y)$(
it can be noted here that since each $\mu _{A}(y),$ $\upsilon _{A}(y)\in $ $%
[0,1]$, a closed and bounded subset of $R$, $\underset{y\in S}{\inf }\mu
_{A}(y)$ and $\underset{y\in S}{\sup }\upsilon _{A}(y)$ exists).Let $z\in S.$
Then $\mu _{A}(x)\leq 
%TCIMACRO{\U{3bc} }%
%BeginExpansion
\mu
%EndExpansion
_{A}(z)$ and $\upsilon _{A}(x)\geq \upsilon _{A}(z)$. Hence 
\begin{equation*}
\left\{ \mu _{A}(x)\vee 
%TCIMACRO{\U{3bc} }%
%BeginExpansion
\mu
%EndExpansion
_{A}(z)\right\} =%
%TCIMACRO{\U{3bc} }%
%BeginExpansion
\mu
%EndExpansion
_{A}(z)..........\ast
\end{equation*}%
and%
\begin{equation*}
\left\{ \upsilon _{A}(x)\wedge \upsilon _{A}(z)\right\} =\upsilon
_{A}(z)..........\ast ^{^{\prime }}
\end{equation*}%
Now$\ <x,\mu _{A}>(z)=\underset{\gamma \in \Gamma }{\inf }%
%TCIMACRO{\U{3bc} }%
%BeginExpansion
\mu
%EndExpansion
_{A}(x\gamma z)$ and $<x,\upsilon _{A}>(z)=\underset{\gamma \in \Gamma }{%
\sup }\upsilon _{A}(x\gamma z).$Since $A=(\mu _{A},\upsilon _{A})$ is an
intuitionistic fuzzy prime ideal of $S,$So $\underset{\gamma \in \Gamma }{%
\inf }%
%TCIMACRO{\U{3bc} }%
%BeginExpansion
\mu
%EndExpansion
_{A}(x\gamma z)=\left\{ \mu _{A}(x)\vee 
%TCIMACRO{\U{3bc} }%
%BeginExpansion
\mu
%EndExpansion
_{A}(z)\right\} $ and $\underset{\gamma \in \Gamma }{\sup }\upsilon
_{A}(x\gamma z)=\left\{ \upsilon _{A}(x)\wedge \upsilon _{A}(z)\right\} .$%
This implies $\underset{\gamma \in \Gamma }{\inf }%
%TCIMACRO{\U{3bc} }%
%BeginExpansion
\mu
%EndExpansion
_{A}(x\gamma z)=%
%TCIMACRO{\U{3bc} }%
%BeginExpansion
\mu
%EndExpansion
_{A}(z)$ and $\underset{\gamma \in \Gamma }{\sup }\upsilon _{A}(x\gamma
z)=\upsilon _{A}(z)$ (using $\ast ,\ast ^{^{\prime }}$).Hence $<x,\mu
_{A}>(z)=%
%TCIMACRO{\U{3bc} }%
%BeginExpansion
\mu
%EndExpansion
_{A}(z)$ and $<x,\upsilon _{A}>(z)=\upsilon _{A}(z).$Consequently,$<x,A>=$ $%
A.$

Conversely, let $A=(\mu _{A},\upsilon _{A})$ be an intuitionistic fuzzy
ideal of $S$ such that $<y,A>=A\ \forall \ y\in S$ with $A(y)$ not maximal
in $A(S)$ and let $x_{1},x_{2}\in S$. Then $A=(\mu _{A},\upsilon _{A})$
being an intuitionistic fuzzy ideal of $S$, $\mu _{A}(x_{1}\gamma x_{2})\geq 
%TCIMACRO{\U{3bc} }%
%BeginExpansion
\mu
%EndExpansion
_{A}(x_{1})$,$\upsilon _{A}(x_{1}\gamma x_{2})\leq \upsilon _{A}(x_{1})$ and 
$%
%TCIMACRO{\U{3bc} }%
%BeginExpansion
\mu
%EndExpansion
_{A}(x_{1}\gamma x_{2})\geq 
%TCIMACRO{\U{3bc} }%
%BeginExpansion
\mu
%EndExpansion
_{A}(x_{2}),\upsilon _{A}(x_{1}\gamma x_{2})\leq \upsilon _{A}(x_{2})$ $%
\forall \gamma $ $\in $ $\Gamma $. This leads to $\underset{\gamma \in
\Gamma }{\inf }%
%TCIMACRO{\U{3bc} }%
%BeginExpansion
\mu
%EndExpansion
_{A}(x_{1}\gamma x_{2})\geq 
%TCIMACRO{\U{3bc} }%
%BeginExpansion
\mu
%EndExpansion
_{A}(x_{1})$ , $\underset{\gamma \in \Gamma }{\sup }\upsilon
_{A}(x_{1}\gamma x_{2})\leq \upsilon _{A}(x_{1}).....\left( \ast \ast
\right) $ and $\underset{\gamma \in \Gamma }{\inf }%
%TCIMACRO{\U{3bc} }%
%BeginExpansion
\mu
%EndExpansion
_{A}(x_{1}\gamma x_{2})\geq 
%TCIMACRO{\U{3bc} }%
%BeginExpansion
\mu
%EndExpansion
_{A}(x_{2})$ , $\underset{\gamma \in \Gamma }{\sup }\upsilon
_{A}(x_{1}\gamma x_{2})\leq \upsilon _{A}(x_{2})......\left( \ast \ast
^{^{\prime }}\right) .$Now two cases may arise viz. Case $(i)$ Either $(\mu
_{A}(x_{1}),\upsilon _{A}(x_{1}))$ or $(\mu _{A}(x_{2}),\upsilon
_{A}(x_{2})) $ is maximal in $A(S)$. Case $(ii)$ Neither $(\mu
_{A}(x_{1}),\upsilon _{A}(x_{1}))$ nor $(\mu _{A}(x_{2}),\upsilon
_{A}(x_{2}))$ is maximal in $A(S).$Case $(i)$ Without loss of generality,
let $(\mu _{A}(x_{1}),\upsilon _{A}(x_{1}))$ be maximal in $A(S)$.Then $%
\underset{\gamma \in \Gamma }{\inf }%
%TCIMACRO{\U{3bc} }%
%BeginExpansion
\mu
%EndExpansion
_{A}(x_{1}\gamma x_{2})\leq 
%TCIMACRO{\U{3bc} }%
%BeginExpansion
\mu
%EndExpansion
_{A}(x_{1})$ , $\underset{\gamma \in \Gamma }{\sup }\upsilon
_{A}(x_{1}\gamma x_{2})\geq \upsilon _{A}(x_{1}).$Consequently $\underset{%
\gamma \in \Gamma }{\text{ }\inf }%
%TCIMACRO{\U{3bc} }%
%BeginExpansion
\mu
%EndExpansion
_{A}(x_{1}\gamma x_{2})=%
%TCIMACRO{\U{3bc} }%
%BeginExpansion
\mu
%EndExpansion
_{A}(x_{1})=\left\{ \mu _{A}(x_{1})\vee 
%TCIMACRO{\U{3bc} }%
%BeginExpansion
\mu
%EndExpansion
_{A}(x_{2})\right\} ,$ $\underset{\gamma \in \Gamma }{\sup }\upsilon
_{A}(x_{1}\gamma x_{2})=\upsilon _{A}(x_{1})=\left\{ \upsilon
_{A}(x_{1})\wedge \upsilon _{A}(x_{2})\right\} .$Case (ii) By the hypothesis 
$<x_{1},A>=A$ i.e $<x_{1},%
%TCIMACRO{\U{3bc} }%
%BeginExpansion
\mu
%EndExpansion
_{A}>=%
%TCIMACRO{\U{3bc} }%
%BeginExpansion
\mu
%EndExpansion
_{A}$ and $<x_{1},\upsilon _{A}>=\upsilon _{A}$ also $<x_{2},%
%TCIMACRO{\U{3bc} }%
%BeginExpansion
\mu
%EndExpansion
_{A}>=%
%TCIMACRO{\U{3bc} }%
%BeginExpansion
\mu
%EndExpansion
_{A}$ and $<x_{2},\upsilon _{A}>=\upsilon _{A}.$ Hence $<x_{1},%
%TCIMACRO{\U{3bc} }%
%BeginExpansion
\mu
%EndExpansion
_{A}>\left( x_{2}\right) =%
%TCIMACRO{\U{3bc} }%
%BeginExpansion
\mu
%EndExpansion
_{A}\left( x_{2}\right) $ and $<x_{1},\upsilon _{A}>\left( x_{2}\right)
=\upsilon _{A}\left( x_{2}\right) \implies \underset{\gamma \in \Gamma }{%
\inf }%
%TCIMACRO{\U{3bc} }%
%BeginExpansion
\mu
%EndExpansion
_{A}(x_{1}\gamma x_{2})=%
%TCIMACRO{\U{3bc} }%
%BeginExpansion
\mu
%EndExpansion
_{A}\left( x_{2}\right) $ and $\underset{\gamma \in \Gamma }{\sup }\upsilon
_{A}(x_{1}\gamma x_{2})=\upsilon _{A}\left( x_{2}\right) .$So $\underset{%
\gamma \in \Gamma }{\inf }%
%TCIMACRO{\U{3bc} }%
%BeginExpansion
\mu
%EndExpansion
_{A}(x_{1}\gamma x_{2})=\left\{ \mu _{A}(x_{1})\vee 
%TCIMACRO{\U{3bc} }%
%BeginExpansion
\mu
%EndExpansion
_{A}(x_{2})\right\} $ and $\underset{\gamma \in \Gamma }{\sup }\upsilon
_{A}(x_{1}\gamma x_{2})=\left\{ \upsilon _{A}(x_{1})\wedge \upsilon
_{A}(x_{2})\right\} $ (using $\left( \ast \ast \right) ,\left( \ast \ast
^{^{\prime }}\right) $).Thus we conclude that $A=(\mu _{A},\upsilon _{A})$
is an intuitionistic fuzzy prime ideal of $S.$
\end{proof}

To end this paper we get the following characterization theorem of a prime
ideal of a $\Gamma $-Semigroups which follows as a corollary to the above
theorem.

\subsection{Corollary}

Let $S$ be a $\Gamma $-Semigroups and $I$ be an ideal of $S$. Then $I$ is
prime iff for $x\in S$ with $x\not\in I$, $<x,X_{I}>=$ $X_{I}$, where $%
X_{I}=\left( \Phi _{I},\Psi _{I}\right) $ is the characteristic function of $%
I$.

\begin{proof}
Let I be a prime ideal of S. Then, by Proposition 2.9, $X_{I}=\left( \Phi
_{I},\Psi _{I}\right) $ is an intuitionistic fuzzy prime ideal of $S$. For $%
x\in S$ such that $x\not\in I$, we have 
\begin{equation*}
\Phi _{I}\left( x\right) =0=\underset{y\in S}{\inf }\Phi _{I}\left( y\right)
\end{equation*}%
and%
\begin{equation*}
\Psi _{I}\left( x\right) =1=\underset{y\in S}{\inf }\Psi _{I}\left( y\right)
\end{equation*}%
Then by Theorem 3.17 $<x,X_{I}>=$ $X_{I}.$

Conversely, let $<x,X_{I}>=$ $X_{I}$ for all $x$ in $S$ with $x\not\in I$%
,Let $y\in S$ be such that $X_{I}(y)$ is not maximal in $X_{I}(S)$. Then $%
\Phi _{I}(y)=0$ and $\Psi _{I}(y)=1$ so $y\not\in I$. So $<y,X_{I}>=$ $X_{I}$%
.So by the Theorem 3.17, $X_{I}$ is an intuitionistic fuzzy prime ideal of $%
S $. So $I$ is rime ideal of S(cf. Proposition 2.9).
\end{proof}

\end{document}